\documentstyle[12pt]{article}  

\def\O{{\cal O}}
\def\Im{\mathop{{\cal I}\!m}}

\begin{document}

\font\fortssbx=cmssbx10 scaled \magstep2
\hbox to \hsize{
\hskip.5in \raise.1in\hbox{\fortssbx University of Wisconsin - Madison}
\hfill\vbox{\hbox{\bf MAD/PH/772}
            \hbox{July 1993}} }

\begin{center}
{\large\bf BLOIS V: SUMMARY TALK\footnotemark}\\[.2in]
Francis Halzen\\
{\it Department of Physics, University of Wisconsin, Madison, WI 53706, USA}
\end{center}

\footnotetext{Talk given at the {\it Vth Blois Workshop on Elastic and
Diffractive Scattering}, June 1993.}

\begin{abstract}

Although we have a theory of strong interactions, its implications for the
physics of average hadron collisions, which is the focus of this series of
conferences, are far from obvious. QCD has nevertheless been a very powerful
guide and I will argue that a consensus description of ``soft" hadronic
physics is emerging at least at the phenomenological level, even though the
message is hidden by a myriad of competing models such as the Lipatov Pomeron,
the jet model, dual topological unitarization, etc. Understanding this physics
is a high priority not only because it is intrinsically interesting (e.g.\ the
search for chiral condensates or for point-like structure of the Pomeron), but
because it dictates the structure of the backgrounds and underlying events in
any hadron collider experiment. In the not so distant future most of particle
physics will rely on such machines and I will therefore emphasize issues
associated with the physics exploitation of future colliders (e.g.\ rapidity
gaps). This is also the first Blois meeting after HERA and the experiments
have already contributed significant novel information on the physics
discussed at this meeting.

\end{abstract}

\thispagestyle{empty}

\vspace{.1in}

\section{Why do total cross sections grow with energy?}

In 1970 we learned that experiments at the Serpukhov accelerator revealed that
the value of the $K^+$-proton total cross section increased with energy. The
hope that this was the crucial experimental hint that would allow us to
understand strong interactions soon faded. Predictions for (at the time)
future accelerators such as the SPS and Fermilab machines were made using fits
with such esoteric parameterizations as imaginary double poles exhibiting our
total inability to learn anything from this exciting discovery. In those
pre-QCD days some ideas emerged which were just awaiting the discovery of this
theory to jump to the forefront. From experimentation\cite{cheng,lipatov} with
QED Feynman graphs it was shown how ladders of electron and photon exchanges
can be summed to yield a cross section which asymptotically rises as
$\log^2\!s$. Others\cite{dcline}, in a
premature discovery of jets, noticed that strange phenomena occurred at the ISR
(large transverse momentum $\pi^0$ production) which seemed to affect particle
production in average, not just rare, cosmic ray events at much higher energy.
They argued that the rise of the total cross section was just the shadow of
the onset of a new mechanism for producing particles. Ever since competing
schools have been discussing rising cross section as either a $t$-channel or a
$s$-channel phenomenon. Both capitalized on the discovery of QCD and claimed
confirmation of their ideas.

Conservatively, one has to conclude that 20 years after the discovery of QCD
we still do not know why total cross sections grow with energy. Extracting the
answer to this question from the lagrangian, which has been quantitatively
verified by an impressive number and variety of experiments, is, in principle,
only a technical problem. It is, unfortunately, not a straightforward
perturbative problem. In the following I will argue that some kind of
qualitative consensus might be emerging despite a wealth of different points
of view on what QCD has to say about hadron collisions. This consensus has
become rather striking at the phenomenological level where I will argue that
everybody is basically constructing the same framework to accommodate the data
with different labels describing almost identical phenomenological structures.

\section{Rising cross section: the $t$-channel point of view}

Determining the asymptotic behavior of very high energy processes in QED is
not a simple Bjorken and Drell problem. The $\O(\alpha^4)$ diagram for
producing electron pairs in photon-photon collisions dominates the
$\O(\alpha^2)$ diagram above a center-of-mass energy $\sqrt s\simeq 1$~GeV;
see Fig.~1a. Anyone who ever built a calorimeter for particle detection knows
that the (inclusive) cross section for pair production by photons in matter is
dominated by the Bethe-Heitler process which grows logarithmically with energy
even though this process is formally higher order than the Compton process.
The relevant diagrams are shown in Fig~1b. Although the basic reason is the
$(\sqrt s)^{-1}$ suppression associated with the fermion propagators in the
leading order diagrams, it is also useful to think of Bethe-Heitler as a
process of order $(\alpha^2)(\alpha\rm\,\log s)$. Large logarithms associated
with pairs can compensate for the additional small coupling constant. The same
argument leads to the well-known result that $\O(\alpha_s^3)$ as well as
$\O(\alpha_s^2)$ diagrams form the leading order QCD contributions to heavy
quark production. In calculating photon-proton elastic scattering the 6th
order process shown in Fig.~1c dominates all lower orders in the very high
energy limit! The examples illustrate how the presence of pairs and loops can
compensate for powers in $\alpha$ when calculating cross sections in the large
$\log~s$ limit.

The laboratory for studying forward proton-proton scattering in QED is the
process $e^+e^- \rightarrow e^+e^-$. It should by now not come as a surprise
that
the first diagram that grows like $\log s$ contains an electron loop and is of
8th order. Each additional loop contributes a power of $\log s$ and in order
to obtain a meaningful result one has to sum (at least) over all the loops
shown in Fig.~2a. One finds\cite{krip}, up to logarithmic factors, that
\begin{equation}
\sum_n \sigma_n \sim s^{J-1}  \label{sum sigma}
\end{equation}
with
\begin{equation}
J-1 = {11\over32} \pi\alpha^2 \,.
\end{equation}
Those of us old enough to know the words realize we found a Regge-behaved
cross section (power-behavior in $s$). The intercept of the Regge pole (the
Pomeron) is $J$  and exceeds unity by an amount $\alpha^2$, reminding us that
the electron loops are the origin of the growth of the cross section with
energy. For $J=1$ the result is indeed energy independent.

\begin{center}
\hspace{0in}

\medskip

\parbox{5.5in}{\small
Fig.~1: Examples of higher order diagrams dominating the high energy behavior
of cross sections.}
\end{center}

A similar result can be obtained in QCD. An exchanged gluon plays the role of
the box as shown in Fig.~2b, and the intercept of the Pomeron therefore
exceeds unity by an amount proportional to $\alpha_s$~\cite{lipatov}
\begin{equation}
J-1 = 12 \, \ln 2{\alpha_s\over\pi} \,. \label{J-1}
\end{equation}
The good news is that we obtained rising cross sections, the bad news is that
power-behaved amplitudes violate unitarity. This is not a problem. Physicists
know a magic recipe to cure any problem of this type: eikonalize.

\begin{center}
\hspace{0in}

{\small
Fig.~2: Ladders of box(gluon exchange) diagrams dominate the QED(QCD) cross
section.}
\end{center}

\section{Road Map to a successful model of forward scattering\break amplitudes}

For the purpose of modifying (\ref{sum sigma}) into a model which satisfies
unitarity we introduce the impact parameter transform of the usual elastic
scattering amplitude $f(s,t)$ for proton-proton scattering
\begin{equation}
a(s,b) = {1\over2\pi} \int d\vec q \, e^{-i\vec q \vec b} \, f(s,t) \,,
\label{a(s,b)}
\end{equation}
where $q^2=-t$, the invariant momentum transfer in the $t$-channel. Unitarity
requires that
\begin{equation}
2\Im a(s,b) = \bigl| a(s,b) \bigr|^2 + a_{\rm in}(s,b) \,. \label{unitarity}
\end{equation}
Here $a_{\rm in}(s,b)$ is the transform of the inelastic amplitude, defined
following the prescription of Eq~(\ref{a(s,b)}). We now introduce a function
$P(s,b)$, the eikonal,
\begin{equation}
a(s,b) = i\left( 1-e^{-{1\over2}P(s,b)} \right) \,,  \label{eikonal}
\end{equation}
with
\begin{equation}
\sigma_{\rm inel} = \int d^2b \left[1-e^{-P}\right] \label{sigma inel}
\end{equation}
and
\begin{equation}
\sigma_{\rm tot} = 2\int d^2b \left[1-e^{-{1\over2}P}\right] \,.
\label{sigma tot}
\end{equation}
The model builder now pretends calculating $P(s,b)$ rather than $a(s,b)$ and
the eikonal construction of Eqs.~(\ref{eikonal})--(\ref{sigma tot})
automatically satisfies the unitarity relation (\ref{unitarity}). So, the
eikonal for the gluon ladder model is of the form
\begin{equation}
P \sim e^{-\mu b} s^{J-1} \,. \label{P}
\end{equation}
We here assumed an exponential impact parameter distribution of whatever the
stuff is that makes up the proton. The spatial extent of the proton in impact
parameter is described by $\mu \simeq 1/R$ which is the radius of the proton.
On a more sophisticated level this impact parameter distribution can be
related to the charge form factor of the proton. Substituting (\ref{P}) in
(\ref{sigma tot}) yields, in the high energy limit, a cross section which is a
black disk with a radius which grows logarithmically with energy
\begin{equation}
\sigma = 2\pi R^2 = 2\pi\left(J-1\over\mu\right)^2  \label{radius}
\log^2\left(s\over s_0\right)
\end{equation}
and
\begin{equation}
R = {J-1\over \mu} \log{s\over s_0}  \,.  \label{R}
\end{equation}
In this limit
\begin{equation}
{\sigma_{\rm el}\over\sigma_{\rm tot}} = {1\over2} \,,
\end{equation}
as can be seen from Eqs.~(\ref{sigma inel}),(\ref{sigma tot}).

The new and improved Pomeron is shown in Fig.~3 where its eikonal structure is
explicitly exhibited. The vertical stacked diagrams are referred to as
exchanges of 1, 2,\dots\ Reggeized gluons. A Reggeized gluon contains further
corrections to the simple ladder discussed previously; see Fig.~3. The
phenomenologist here finds a grab-bag of models. Although a model
parameterizing the full structure shown in Fig.~3 was constructed by
Kopeliovich et al., the KNP model\cite{krip}, others isolate subsets of
diagrams in order to model forward scattering. In elastic and diffractive
scattering the protons retain their identity, therefore no quantum numbers are
exchanged in the $t$-channel. That is the definition of the Pomeron, it must
therefore have positive charge conjugation and naively corresponds to the
exchange of at least 2 gluons. Such a diagram has pathologies at $t=0$,
which can, however, be eliminated by the use of gluon propagators which are
regularized at $k^2=0$. Such a propagator has, for example, been constructed
by Cornwall\cite{ducati}.

Lipatov\cite{lipatov} introduces the ``reggeized gluon\rlap", which is
essentially gluon exchange with Sudakov suppression. Ladders of two reggeized
gluons, the $P$ term in Fig.~3, form the Lipatov Pomeron. Clearly, ladders of
three reggeized gluons represent the odderon, an odd charge conjugation
contribution to forward scattering which can differentiate the high energy
behavior of $pp$ and $\bar pp$ forward scattering amplitudes. From the triple
reggeized gluon exchange diagram Gauron obtains\cite{gauron}
\begin{equation}
(J_{\rm odd}-1) \ge 0.13 (J-1) \,.
\end{equation}

\begin{center}
\hspace{0in}

{\small
Fig.~3: Structure of the perturbative Pomeron?}
\end{center}

Finally, it has been argued on phenomenological grounds that Pomeron exchange
is a simple factorizable structure like photon exchange --- just like a photon,
but with opposite charge conjugation. This is the Landshoff-Donnachie model,
which has had impressive phenomenological successes in not just accommodating
data, but in actually making predictions\cite{cudell}.

\section{Forward scattering as the shadow of particle production}

The unitarity relation (\ref{unitarity}) reminds us that elastic scattering
cannot be understood without understanding inelastic scattering. Elastic
scattering is the shadow of particle production and, as previously mentioned,
several features of accelerator as well as cosmic ray data suggest that jet
production affects the behavior of average interactions at very high energy. At
low energies, say $\sqrt{s} < 0.1$\,TeV, particle production in hadron
collisions exhibits Feynman scaling, KNO scaling holds and transverse momenta
are limited. Events with jets are very rare, but QCD predicts that their
frequency increases rapidly with energy. At a c.m.\ energy of order 1\,TeV the
production of jets is no longer a rare phenomenon with
\begin{equation}
\sigma_{\rm jet} \, (p_T>3\rm\,GeV) \simeq 20\,mb \;.
\end{equation}
Jet production begins to affect the properties of minimum-bias events. Feynman
scaling is violated, mostly in the central region where jet fragmentation
contributes a new source of secondaries. As jets with low $p_T$-values are
more frequent, the additional pions are soft and populate the central region.
In the KNO-distribution a high multiplicity tail associated with high
multiplicity jet events becomes apparent. The $\left<p_T\right>$-value will
grow with energy. Most dramatically, as high $p_T$ jet events also have a high
multiplicity, a correlation between transverse momentum and multiplicity
develops. Contrary to what is expected on the basis of pure kinematics, high
$p_T$ events are, on average, also high multiplicity events. This correlation,
first observed by the Japan-Brazil collaboration in cosmic ray emulsion
chambers, has been confirmed by collider experiments\cite{blocketal}.

Evidence on particle interactions at the very highest energies, in fact in or
near the SSC energy range, has been collected in exposures of large, high
altitude emulsion chambers to the cosmic ray beam. The most prominent feature
of these events is that a large fraction contains ``halos" which appear as
spots of extremely high optical density on the X-ray film\cite{feinberg}.
A most intriguing
feature is that the halos in each event are aligned to a remarkable degree.
This bizarre observation might have a most conventional explanation.
Halos are electromagnetic showers produced by the most energetic $\pi^0$ in
jets. In a two-jet event the produced jets and the projectile/target fragments
form, by momentum conservation, a plane which will intersect the emulsion thus
defining a line along which the fragmentation products of both jets as well as
those from the colliding hadrons will line up. This picture naturally explains
the alignment of 3~halos. In accelerator language events with 4~halos are
three-jet events. It is a fact of QCD that also the additional jet is
preferentially produced in the plane of the first two. This leads to the
alignment of 4~halos. An accelerator physicist would classify these events as
mini-jet events. If in a roughly 20\% subsample of events jets are visible to
the naked eye, it is reasonable to assume that most high energy events are jet
events!

Because relatively soft partons are the source of the jet phenomena described
above, gluons rather than quarks will dominate particle production at very
high energy. The jet cross section is symbolically given by
\begin{equation}
\sigma_{\rm jet} = \int\limits_{4P^2_{\rm T\rm\,min} / s} dx \, dx' \, g(x) \,
g(x') \sigma(gg\to \rm jets) \,,  \label{sigma jet}
\end{equation}
with the gluon structure function $g(x)$ of the form
\begin{equation}
g(x) \sim {1\over x^J} (1-x)^n \,.
\end{equation}
The parameter $J$ controls the shape of the bremsstrahlung spectrum
determining the number of soft gluons with $g(x) \sim 1/x^J$. The steep soft
gluon spectrum controls the behavior of the integral of Eq.~(\ref{sigma jet})
in the high energy limit. In the limit where $\sigma(gg\to \rm jets)$ varies
slowly with energy, we obtain
\begin{equation}
\sigma_{\rm jet} \sim \left(s\over s_0\right)^{J-1}  \label{HE limit}
\end{equation}
and $s_0=4p_{T\rm\,min}^2$, a result reminiscent of Eq.~(\ref{sum sigma}). The
impact parameter treatment of the gluons will make the connection between jet
production and the total cross section via the unitarity relation
(\ref{unitarity})
\begin{equation}
P = e^{-\mu b} \sigma_{\rm jet} \,. \label{P2}
\end{equation}
The result of Eqs.~(\ref{HE limit})--(\ref{P2}) is identical to the one
previoulsy obtained in Eq.~(\ref{P}) and we recover the same expression for the
total cross section, i.e.\ Eqs.~(\ref{radius}),(\ref{R}), previously obtained
by summing gluon exchanges in the $t$-channel. In the $s$-channel picture the
large number of gluons, not large jet cross sections, is the origin of large
total cross sections at high energy. In deriving (\ref{HE limit}) we indeed
assumed that $\sigma(gg \rightarrow \rm jets)$ is weakly varying with
energy. State of the art $s$-channel models treat ``partons in hadrons" using
the same optical models developed to describe ``nucleons in nuclei\rlap.''
Here, as before, the black disk behavior saturates the Froissart bound. The
limiting size $m_\pi^{-2} = 60$\,mb in the old Froissart bound is here
replaced by $[(J-1)/\mu]^2 = 0.05$\,mb for $\mu^{-1} = R_N = 0.7$\,fm and $J-1
\simeq 0.1$ as required by actual fits to total cross section data\cite{block}.
Note that this is not quite compatible with~(\ref{J-1}).

Notice that we introduced a cutoff $s_0$ or $p_{T\rm\,min}$ regulating the soft
divergence of the jet cross section. Its physical origin is clear. Jets are
produced by the exchange of a colored gluon; see Fig.~4. $p_{T\rm\,min}$
describes the transverse momentum for which its wavelength $\lambda \sim
1/p_{T\rm\,min}$ is such that it begins to ``see" the whole colorless proton
rather than its colored constituents. Jet production is switched off.
Therefore $p_{T\rm\,min}$ is inversely proportional to the size of the
interacting hadrons
\begin{equation}
p_{T\rm\,min} \sim {1\over R} \,. \label{pTmin}
\end{equation}
This relation implies that the rise of the total cross section for different
beams and targets are related. We will return to this further on.

\begin{center}
\hspace{0in}

{\small
Fig.~4: Jet production by gluon exchange.}
\end{center}

\section{Intermezzo: are we all doing the same thing?}

The phenomenological textures used by various groups to describe either high
energy event structures or forward scattering amplitudes are indeed alarmingly
similar. The words describing them differ. This is graphically shown in
Fig.~5. Everybody has a basic building block: a Reggeized gluon\cite{maor},
a string\cite{kanki}, or a hard collision producing jets\cite{block}.
These are summed to give a power-behaved
amplitude which is subsequently associated with an eikonal in order to
guarantee unitarity. The diversity is not all that surprising --- if you asked
the same phenomenologists how hadrons are produced in $e^+e^-$ interactions
they might use different words, as illustrated in Fig.~5, even though in this
case everybody agrees on what is going on.

\begin{center}
\hspace{0in}

\parbox{5.5in}{\small
Fig.~5: Nomenclature for hadron production in $e^+e^-$ and hadron collisions:
quarks and gluons, strings and jets.}
\end{center}

This is not just a cute remark. As a case in point one can take the
phenomenological analyses of forward amplitudes by Kopeliovich et
al.\cite{krip} and Block et al.\cite{block}, $t$-channel and $s$-channel
``inspired" QCD respectively, and
construct a translation table of the fitted parameters. This does not mean
that the modeller cannot get into trouble. The string model does, a priori,
not have the $p_T$-effects that inspire the jet model. They were subsequently
incorporated by allowing hard scattering at the vertices linking the hadrons
and the strings. Now the difference with the jet model becomes more subtle
because the string can be interpreted as a fragmenting jet; see Fig.~6.

\begin{center}
\hspace{0in}

{\small
Fig.~6: String or fragmenting jet?}
\end{center}

Although all these analyses are QCD-inspired it is fair to say that the models
are more ``inspired" than ``QCD"\cite{QCD}. Even at the level of perturbation
theory, the
structure of Fig.~5 has not been rigorously derived from QCD. Even if such a
derivation could be produced, there is no guarantee that the picture is not
essentially modified by infrared effects\cite{white}. Given the lack of
theoretical rigor, the critical question is whether the phenomenology is giving
us confidence that we are on the right track.

\section{If we are all doing the same thing, is it correct?}

The total cross section does rise, maybe even as $\log^2\!s$ although there
is no hard evidence for that\cite{valin}. This debate will undoubtedly resume
with the announcement at this conference that CDF measures\cite{albrow} a total
cross section of $80.6\pm2.3$~mb, in excess of the 1800~GeV value suggested by
previous measurements shown in Fig.~7a. Figures~7a,b,c show how the jet
model\cite{krisch}
adequately describes forward scattering. All schools can produce similar fits.
These do not represent solid evidence for the asymptotic predictions of
Eqs.~(\ref{radius}),(\ref{R}). The asymptotic gluon term, previously
discussed, contributes less than half of the total cross section at the
highest collider energy. In order to fit the data it is essential to include
``low energy" contributions which introduce a lot of ambiguity and extra
parameters and can be exploited to accommodate the data.

\newpage

\begin{center}
\hspace{0in}

\parbox{5.5in}{\small
Fig.~7: Fits to the total cross section, $\rho$-parameter, and forward slope of
the differential cross section in $pp$ and $p \bar p$ interactions; from
reference\cite{block}.}
\end{center}
\eject

Maybe the most tangible hint for the approach to a black disk behavior is the
evidence that the second derivative of the forward slope, which is predicted
to be negative, is changing sign near 1800~GeV\cite{block}. At lower energies
this curvature is indeed positive.

The $\rho$ parameter at 540~GeV is no longer a problem. The remeasured value
is consistent with expectations as can be seen from Fig.~7b. The excitement of
the last Blois meeting is gone. The models' credibility should not be raised
by the fact that they could not accommodate the previous large value. They were
just prevented from doing so by the measurements of the total cross section at
1800~GeV and analyticity. Eddington warned us not to take an experimental
result seriously until it is confirmed by theory. Also, the difference between
the old and new value is only $2.3 \sigma$ for an experiment with incredibly
challenging systematic problems. In gamma ray astronomy $4 \sigma$ effects are
routinely ignored!

Is there an odderon? In perturbative QCD there is\cite{gauron}, but arguments
were presented that it may disappear after infrared effects are taken into
account\cite{white}.
A precise measurement of the difference between the $pp$ and $p\bar p$ total
cross sections seems not to be in our future. Our only hope is to investigate
the dip structure near $t=-1\rm~GeV^2$ in the differential cross section.
Destructive interference of Pomeron and double Pomeron exchange is claimed to
be the origin of the minimum in the cross section; see Fig.~8. Now we are
told, however, that at high energy there is a dip in $pp$, but not in $p\bar
p$. So we need a new ingredient to fix this and the odderon just fits because
it contributes differently to the two reactions. Interestingly, we heard at
this conference that rather large values of the elastic polarization, which we
tend to associate with non-asymptotic effects, do not decrease with energy
near the structure at $t=-1\rm~GeV^2$~\cite{krisch}. This is possibly another
hint for the presence of the odderon.

\begin{center}
\hspace{0in}

\parbox{5.5in}{\small
Fig.~8: Three contributions are needed to accommodate the dip structure in $pp$
and $p\bar p$ elastic scattering.}
\end{center}

Measuring cross sections with different beam and target particles has been a
powerful phenomenological tool in the past. It provided us with early evidence
for the quark structure of hadrons via the additive quark model. It is clear
that in most, but maybe not all, models the additive quark model is at best
approximate and breaks down at the higher energies where gluons dictate the
interactions\cite{martin}. HERA has presented us with the opportunity to
perform such a test\cite{durand} by dramatically increasing the energy-reach of
the data on the total photoproduction cross section\cite{albrow}. We will here
just treat the photon as a
$\rho$-meson with 2 quark constituents. Given this additional assumption the
photoproduction cross section can be calculated from $pp$ data in the jet
model. The low energy contributions to both cross sections, which are of quark
origin, are related by a factor 3/2 given by the additive quark model. The
rise of the total cross section, which is the shadow of jet production by
gluons, can be related using Eq.~(\ref{pTmin})
\begin{equation}
{ \left( p_{T\rm\,min}^2 \right)_{\gamma p} \over
  \left( p_{T\rm\,min}^2 \right)_{pp} } = {3\over2} \,.
\end{equation}
The result of this straightforward exercise\cite{fletcher} is shown in Fig.~9.
It works.

\begin{center}
\hspace{0in}

\parbox{5.5in}{\small
Fig.~9:  The jet model relates the high energy behavior of photoproduction with
that of total cross sections.}
\end{center}

At this conference several new measurements on diffraction dissociation were
presented. They are sensitive to the critical parameter $J$ first introduced in
Eq.~(\ref{sum sigma}) which can be determined by measuring both the dependence
of the cross section on energy and on the produced mass $M$ via\cite{frankfurt}
\begin{equation}
s {d\sigma\over dt\,dM^2} \sim e^{bt} \left(s\over M^2\right)^{2(J-1)} \,.
\end{equation}
The CDF experiment finds\cite{albrow} a value which nicely matches a value in
the vicinity of 0.1 obtained from phenomenological analysis of forward
scattering
\begin{equation}
J-1 = 0.115 \pm 0.013 \,.
\end{equation}

\section{Perturbative QCD, elastic scattering and\hfill\break
 color transparency}

A series of talks emphasized progress in putting the QCD counting rules
for large angle elastic scattering on a solid theoretical foundation. That the
elastic scattering cross section of two protons at right angles falls with
energy as a power of energy $s^{-10}$ is one of the most unusual and best
tested predictions of QCD\cite{islam}. The data do, however, show some
oscillatory behavior around the power behavior which is associated with soft
gluon corrections. Soft gluons are large objects with large geometric cross
sections. Hard gluons, on the contrary, are small objects. In doing the large
angle elastic scattering experiment on a nuclear target, the nucleus can be
thought of as a brick wall filtering out the soft gluons thus supplying a
purified view of the true perturbative prediction. The first data in support
of this prediction were hotly debated.

\section{Ode to Hera}

A summary talk at this conference must highlight the impressive
contributions the ZEUS and H1 experiments at HERA are now making to our
endeavors to understand strong interactions. I already discussed the
measurement of the photoproduction cross section. The cross section is
asymptotically dominated by the gluon structure of the photon. Such photonic
gluons produce the jets which drive the rise of the cross section in the model
discussed; see Fig.~10. Interestingly, the experimental identification of this
structure function, reported by both experiments at this conference, is really
our first glimpse at the physics that dominates photoproduction at high
energy. Indirect evidence for the gluon structure of photons had previously
been observed in photon-photon interactions at TRISTAN\cite{amy}. That the high
energy proton is a gluonic structure can be best illustrated\cite{levin}
by the statement that the new measurements of small-$x$ structure functions
imply that the packing factor of gluons inside protons is the same as that for
nucleons in iron at $x \sim 10^{-4}$ and $Q^2 \sim 10\rm~GeV^2$.

HERA rediscovered, probably to the surprise of the younger generation, that
the Pomeron exists. They observe a sample of events with a
large rapidity gap between the electroproduced hadrons and the proton; see
Fig.~11a. Jet production in such events will be studied in future
higher luminosity runs thus allowing us to probe the partonic structure of
the Pomeron. We were reminded\cite{schlein} at this conference of the evidence
for a hard partonic component of the Pomeron revealed by the production of jets
in Pomeron-proton collisions at the $Sp \bar pS$; see Fig.~11b. So, HERA might
reveal the gluonic structure of the Pomeron next year just like it
demonstrated its presence in the photon this year.

\begin{center}
\hspace{0in}

{\small
Fig.~10: Photons resolved into gluons dictate the high energy photoproduction
of hadrons.}
\end{center}

\begin{center}
\hspace{0in}

{\small
Fig.~11: Hadron production by Pomerons in electroproduction and hadron
interactions.}
\end{center}

The importance of this study cannot be overstated. The experimental proof of
the existence of a structure function relies on two of its properties:
factorization and momentum conservation (the momentum sum rule must be
satisfied). The Pomeron, which is in its simplest form a 2-gluon structure,
most likely satisfies neither one. So, it is not clear how to proceed. Cynics
will say that the Pomeron structure function revealed by the $Sp \bar pS$
experiment is no more than a remnant of the cuts on the data. The origin of
its hard structure, with 30\% of the events requiring a $\delta(1-x)$
component, is less than obvious. There are more questions than answers here and
HERA will hopefully bring us yet another pleasant surprise.

\section{Why are we doing this?}

It is fair to say that the question of why the total cross section rises with
energy is not as glamorous as it once was. Are the problems we study out of
the mainstream of particle physics? Definitely not, what is, in fact, more
mainstream than the question of how to find the Higgs. The recent experimental
as well as theoretical activity surrounding rapidity gaps as a signature for
the Higgs particle at future colliders has brought ``soft physics" back to
center stage; more about that later. What about Higgs production by Pomerons?
The diagram for producing a Higgs in association with two rapidity gaps is
shown in Fig.~12a. The pair of gluons is in a color singlet, the rapidity gap
is allegedly produced because no color is exchanged, and therefore one can
also visualize the diagram as in Fig.~12b, i.e.\ Higgs production by Pomerons.
The diagram in Fig.~12a looks like a $(\alpha_s^2)$ QCD correction to Higgs
production by gluon fusion and contributes indeed at the 1\% level. The
calculation\cite{bialas} only evaluates the elastic part of the cross section,
i.e.\ the protons do not break up, and the inclusive cross section may be
larger. One might actually question whether the diagram in Fig.~12a really
represents the dominant contribution to Higgs production.

\begin{center}
\hspace{0in}

{\small
Fig.~12:  Higgs production by Pomerons.}
\end{center}

One should not rule out the possibility of making a fundamental discovery. At
this conference we heard many suggestions ranging from the discovery of the
breakdown of forward dispersion relations, possibly associated with additional
dimensions in a string theory\cite{khuri}, to the observation of chiral
condensates\cite{taylor}. Chiral
condensates are not just an interesting theoretical idea, they may be the
explanation for some puzzling phenomena observed in cosmic ray experiments.
This includes Centauro events. Let me remind you that the QCD lagrangian for
massless quarks is invariant under global isospin and global chiral
transformations. The symmetry is spontaneously broken leading to the
interpretation of pions as massless Goldstone bosons. As a result quark
masses are generated. The possibility cannot be excluded that in a high energy
collision a macroscopic, localized region of spacetime exists where chiral
symmetry is restored. Imagine indeed that in a head-on collision of protons a
significant fraction of the incident momentum is thermalized leading to a hot
spot which is in the chirally symmetric phase. This region cools while the
hadronic debris expands outward and, when the hot spot drops below the critical
temperature, it undergoes a phase transition to the broken phase. This vacuum
inside the fireball may be disoriented from its external or ambient value
as shown in Fig.~13a. This orientation may be thought of as a vector in
isospin space as shown in Fig~13b. In the late stages of the interaction the
two vacua will meet and the inside vacuum will relax or adjust to the
orientation of the ambient one by emitting Goldstone bosons, i.e.\ soft pions.
Depending on the mismatch this adjustment will require the emission of mostly
charged or mostly neutral soft pions. The production of a chiral condensate
can therefore result in the population of a region of phase space with mostly
neutral(charged) pions. At this conference we were shown\cite{iwai} an event
containing a cluster of 32 photons (from $\pi^0$ decay) with only one
accompanying charged particle. Copious production of chiral condensates will
result in anomalously large charge fluctuations. The famous
Centauro/mini-Centauro events\cite{hasegawa} can be interpreted just this way
and may be a signature of chiral condensates.

\begin{center}
\hspace{0in}

\parbox{5.5in}{\small
Fig.~13: Different orientations of the QCD vacua inside and outside a fireball.
The orientation represents a direction in isospin space.}
\end{center}

Finally the physics of the Blois workshop is important because future particle
physics will be mostly done with hadron colliders. We have learned from the
operation of the present generation of colliders that, whatever the physics
under study, in the end some Monte Carlo generator is needed to deal with
backgrounds or the effects of the underlying event. I have the impression few
still realize that PYTHIA, HERWIG, SYBIL and the like were not given by God
but are based on the physics discussed at this conference. We have actually
come to the rather undesirable situation that the jet model is used in most
Monte Carlo's. An in-depth study of these ideas, as performed at this
conference, is critical and although it might be considered
engineering by some, it is essential. SSC and LHC cannot avoid facing the
issues we debated. The value of the total cross section might, in fact, just be
the most important physics parameter of the SSC/LHC colliders. This again is
one of these numbers on which almost everybody agrees: 120~mb for the SSC,
105~mb for the LHC. But then remember that we are all doing essentially the
same fits\dots

\section{PS: Rapidity gaps}

The flurry of activity since Blois IV on the use of rapidity gaps as a
signature for electroweak processes provides us with yet another example of
the urgency of understanding minimum-bias physics. The subject deserves all
the attention it has attracted\cite{duca}. The hope has been raised that
rapidity gaps
will present us with a much needed opportunity for sharpening the experimental
signature of Higgs production in hadron collisions. In Fig.~14 we contrast an
event where a Higgs particle is produced by a pair of W's with a minimum bias
event. The different color flow in the events will result in different
rapidity distributions of the produced hadrons. The simplest way to visualize
this flow in a minimum-bias event is to think of protons as quark-diquark
systems. The quark triplet color in one hadron is bleached by the antitriplet
color of the diquark in the other hadron and vice versa. Two strings are
stretched between the colliding particles and their fragmentation uniformly
populates the rapidity distribution; see Fig.~14. Events where colorless
electroweak particles are exchanged are essentially different in the sense
that all color flow is localized near the beam particles, no color is
exchanged leaving a gap in rapidity space with no particles (except for the
fragments of the decaying Higgs) between the beam particles.
Also, jets appear in the fragmentation regions as the result of the recoil of
the quarks emitting the weak bosons. Rapidity gaps, well-known to this
audience as a signature for colorless Pomeron exchange, are here a signature
for Higgs or electroweak exchange in general.

Is this too good to be true? It is, and here a series of very hard questions
must be raised whose answers are intimately connected with the physics
discussed at this meeting. How often do real gaps disappear? They can be
filled by (coherent) radiation of particles by the active quarks that emitted
the $W$'s. Alternatively, particles associated with the underlying event, e.g.\
additional jet production, can fill the gap. This is referred to as the
survival probability. In the jet model it is readily computed to be $\exp(-P)$
with $P \sim \sigma_{jet}$, i.e.\ the probability that no jet production
occurs.
In Fig.~15 we show the rapidity of hadrons in a sample of Higgs events.
Also shown is the same distribution with the underlying event removed.

\begin{center}
\hspace{0in}

{\small
Fig.~14: Strings fragmenting into hadrons in minimum-bias and Higgs events.}
\end{center}

How often do fake gaps appear? Fluctuations in regular events can produce fake
gaps. One might expect that the probability is exponentially suppressed, but
in a jet model this is not the case. Fluctuations associated with multiple jet
production are more likely. Also, are gap events, which are in no way
associated with electroweak phenomena, produced by the exchange of a pair of
gluons in a color singlet state? The answer to this question is far from
obvious because the fact that the net color is zero does not imply that the
gluons cannot radiate into the gap. The subject has obviously raised more
questions than it has answered and in a concerted effort experimentalists and
theorists are tackling them. The experimentalists are using jet production by
photon exchange as a laboratory for studying rapidity gap physics. Preliminary
data indicate\cite{albrow} the observation of gaps. The $10^{-2}$ to $10^{-3}$
observation rate is consistent with estimates that gaps are produced by 2-gluon
exchange at the 10\% level and that such gaps have a survival probability of
10\%. This is a challenging avenue of research which is likely to intensify in
the future.

\begin{center}
\hspace{0in}

\parbox{5.5in}{\small
Fig.~15: Rapidity distribution of hadrons in events where a Higgs has been
produced, including(solid line) and excluding(dashed line) hadrons associated
with the underlying event, i.e.\ mini-jet production in the case of PYTHIA.}
\end{center}

\section*{Acknowledgements}

I thank Jean Ren\'e Cudell for his patience in answering my questions. This was
a very exciting meeting and I feel that H.~Fried, K.~Kang, and C.~I.~Tan had a
lot to do with that. This research was supported in part by the U.S.~Department
of Energy under Contract No.~DE-AC02-76ER00881, in part by
the Texas National Research Laboratory Commission under Grant No.~RGFY93-221,
and in part by the University of Wisconsin Research Committee with funds
granted by the Wisconsin Alumni Research Foundation.

\end{document}